\documentclass[aps,reprint]{revtex4-2}
\usepackage{mathrsfs}
\usepackage{amsmath}
\usepackage{gensymb}
\usepackage{amsfonts}
\usepackage{amssymb}
\usepackage{amsthm}
\usepackage{graphicx}
\usepackage{natbib}
\usepackage{xcolor}
\usepackage{hyperref}
\usepackage{bm}
\usepackage[caption=false]{subfig}
\usepackage{verbatim}
\usepackage{siunitx}
\usepackage{tabularx}
\usepackage{booktabs}
\usepackage{multirow}
\usepackage{mathtools}

\begin{document}

\title{Nature of the 175–180 cm$^{-1}$ Raman Feature in Altermagnetic $\alpha$-MnTe}
\author{ Pankaj Bhardwaj$^{1,\dagger}$, Rajib Sarkar$^{1,\dagger}$,  Naresh Shyaga$^1$, Subhransu Kumar Negi$^1$, Kartick Biswas$^1$, Susmita Jana$^{2,3}$,  Pavan Nukala$^1$, B. R. K. Nanda$^{2,3}$ and Dhavala Suri$^{1,}$}

\email{dsuri@iisc.ac.in}
\thanks{$^{\dagger}$These authors contributed equally.}

\affiliation{$^1$Centre for Nanoscience and Engineering, Indian Institute of Science, Bengaluru, Karnataka 560012, India} 
\affiliation{$^2$Condensed Matter Theory and Computational Lab, Department of Physics, IIT Madras, Chennai, 600036, Tamil Nadu, India}
\affiliation{$^3$Center for Atomistic Modelling and Materials Design, IIT Madras, Chennai, 600036, Tamil Nadu, India}

\begin{abstract}
The 175~cm$^{-1}$ Raman mode in $\alpha$-MnTe has long been assigned to the material's only symmetry-allowed phonon ($E_{2g}$), yet first-principles calculations consistently place this mode below 100~cm$^{-1}$. Recent works have offered three competing explanations for the discrepancy: a weak symmetry-lowering ($D_{6h}\rightarrow D_{3h}$) phonon, an electronic plasmon arising from intrinsic hole self-doping, or an extrinsic MnTe$_2$ impurity signature. Here we resolve this controversy using stoichiometry-controlled molecular beam epitaxy (MBE). We deliberately vary the fraction of pyrite-type MnTe$_2$ secondary phase relative to the $\alpha$-MnTe matrix. Across this series, the 175~cm$^{-1}$ mode tracks the presence of MnTe$_2$ essentially one-to-one -- present wherever MnTe$_2$ is detectable, absent in stoichiometric, single-phase $\alpha$-MnTe at nominal LASER powers. When LASER power is ramped up, an additional mode emerges in MnTe as well positioned close to $\approx$ 175~cm$^{-1}$; our detailed temperatures dependendent Raman analysis identifies the LASER power scorched region close to a distorted MnTe phase, confirmed via transmission electron microscopy.  This direct, growth-controlled correlation  identifies the mode as an extrinsic MnTe$_2$ impurity signature or as a distorted MnTe feature, rather than an intrinsic phonon or plasmon feature, providing a practical diagnostic for phase purity in MBE-grown $\alpha$-MnTe and clarifying which  Raman features can be reliably attributed to the intrinsic altermagnetic phase.
\end{abstract}

\maketitle

Altermagnetism has recently been established as a distinct, third class of collinear magnetic order, standing apart from conventional ferromagnetism and antiferromagnetism~\cite{Smejkal2022PRX1,Smejkal2022PRX2}. In a conventional collinear antiferromagnet, the two spin sublattices are related by translation (as in the N\'eel state of NiO) or by inversion/time-reversal combined with a proper rotation, enforcing spin-degenerate bands throughout the Brillouin~\cite{Yuan_2021}. Altermagnets instead host sublattices connected by a proper rotation alone (or a rotation combined with a nonsymmorphic translation), with no accompanying translation or inversion~\cite{Fender_2025}.

Among the growing family of altermagnetic candidates, $\alpha$-MnTe has emerged as the archetypal and most intensively studied material. It crystallizes in the hexagonal NiAs-type structure (space group $P6_3/mmc$, point group $D_{6h}$), with Mn$^{2+}$ spins ferromagnetically aligned within the basal $ab$-plane and antiferromagnetically stacked along $c$, ordering below $T_N \approx 307$~K \cite{Wu2025,Zhang_2026}. The two Mn sublattices are related by a nonsymmorphic six-fold screw rotation rather than by inversion or translation, placing MnTe in the $g$-wave altermagnetic class. Its appeal lies in a combination of structural and chemical simplicity -- a binary compound long studied as a prototypical antiferromagnetic semiconductor and as the parent phase of Cd$_{1-x}$Mn$_x$Te-type dilute magnetic semiconductors -- together with a wide band gap ($E_g \approx 1.3$--1.4~eV) that makes it experimentally accessible~\cite{StachowWjcik_2000}.

\begin{figure*}[!ht]
\centering
\includegraphics[width=18cm]{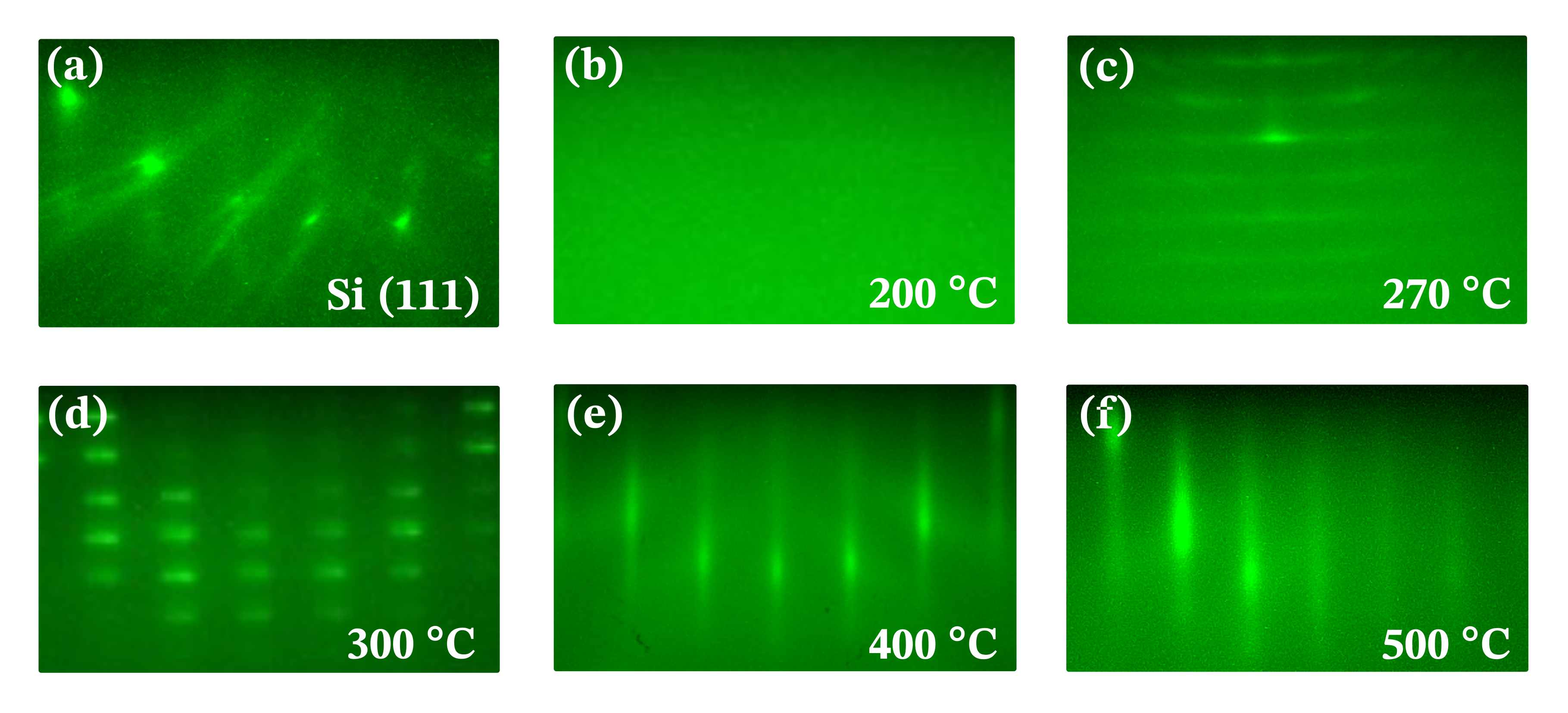} 
\caption{ RHEED patterns of (a) the bare Si(111) substrate and of films grown at (b) \SI{200}{\celsius}, (c) \SI{270}{\celsius}, (d) \SI{300}{\celsius}, (e) \SI{400}{\celsius}, and (f) \SI{500}{\celsius}.
} 
\label{RHEED}
\end{figure*}

This combination has enabled MnTe to anchor essentially every major experimental confirmation of altermagnetism to date: direct observation of the predicted $g$-wave band splitting by spin- and angle-resolved photoemission~\cite{Krempasky2024,Lee2024,Osumi2024}; a giant X-ray magnetic circular dichroism signal in a material with zero net moment~\cite{Hariki2024}; a magnetic-field-controllable anomalous Hall effect~\cite{Sarkar_2026,Kluczyk2024}; nanoscale imaging and electrical switching of N\'eel domains~\cite{Amin2024}; and direct evidence for altermagnetically split, chiral magnon branches by inelastic neutron scattering and circular-dichroism~\cite{Liu2024ChiralSplitMagnon,Jost2025}. MnTe has consequently become the most heavily studied altermagnetic material, with dozens of papers appearing within 2025 alone~\cite{Thapa2026,Negi2025}, and now serves as the reference system against which every new altermagnetic probe -- transport, optical, or spectroscopic -- is first calibrated.

Raman spectroscopy has played a comparatively modest but persistent role in this effort, and the reported mode structure of $\alpha$-MnTe has long been treated as settled. Group theory for the $D_{6h}$ structure predicts nine zone-center optical degrees of freedom from the two Mn and two Te atoms per unit cell, decomposing into one Raman-active $E_{2g}$ mode, one silent $B_{1u}$ mode, two infrared-active $A_{2u}$ and two $E_{1u}$ modes, one silent $B_{2g}$ mode, and one silent $E_{2u}$ mode~\cite{Wu2025} -- so that $E_{2g}$ is, by symmetry, the only phonon expected in the Raman spectrum. Since the earliest Raman study of MnTe by Mobasser and Hart~\cite{MobasserHart1985}, which reported a phonon near 178~cm$^{-1}$ alongside a broad two-magnon feature near 270~cm$^{-1}$, this $\sim$175--180~cm$^{-1}$ line has anchored essentially every subsequent Raman study of the material, in bulk crystals, epitaxial films, and exfoliated flakes alike~\cite{Szuszkiewicz1997,Zhang2020,Kluczyk2024,Bossini2021}. It has routinely been assigned to the $E_{2g}$ phonon on the strength of this symmetry argument alone, and used as a reference feature for tracking spin-phonon coupling through $T_N$, identifying coherent-phonon oscillations in ultrafast pump-probe experiments, and, more recently, as a benchmark against which additional low-frequency features -- one- and two-magnon excitations, and modes near 120 and 140~cm$^{-1}$ tentatively attributed to a Te impurity phase~\cite{Szuszkiewicz2014,azam2026} -- are calibrated.

This assignment sits uneasily with theory: first-principles phonon calculations for the $D_{6h}$ structure consistently place the $E_{2g}$ mode well below 100~cm$^{-1}$ (83--91~cm$^{-1}$, depending on functional and Hubbard $U$~\cite{Wu2025,Thapa2026}) -- a near factor-of-two discrepancy that is unusual for DFT phonon calculations in a structurally simple binary compound, and one that went largely unaddressed until recently.

This discrepancy has, over the past year, turned the 175~cm$^{-1}$ mode from a routine reference peak into an actively contested problem. Polarization-resolved Raman measurements show that the mode's selection rules are inconsistent with an $E_{2g}$ assignment: the peak is strong in co-polarized channels but vanishes in cross-polarized geometries, whereas an $E_{2g}$ phonon should scatter comparably in both~\cite{Wu2025,Palasyuk2026}. This, together with second-harmonic-generation polarimetry indicating broken inversion symmetry, motivated a proposed symmetry-lowering distortion from $D_{6h}$ to $D_{3h}$, converting the nominally silent $B_{1u}$ mode into a Raman-allowed $A_1'$ mode calculated at $\sim$177~cm$^{-1}$, apparently matching both frequency and polarization behavior~\cite{Wu2025}. This ``symmetry-leakage'' hypothesis has since been challenged: fully relaxed DFT+U optimizations, checked against multiple functionals and a wide range of $U$, converge back to the centrosymmetric $P6_3/mmc$ structure with zero residual distortion. A Placzek-formalism calculation of the distortion-induced mode's Raman activity -- scaling as the square of the vanishingly small ($\delta \sim 0.1\%~c$) symmetry-breaking displacement -- finds an intensity nearly two orders of magnitude too weak to explain the observed peak~\cite{Thapa2026}. Ruling out both the $E_{2g}$ assignment and the symmetry-leakage mechanism, the same work instead proposes that the 175~cm$^{-1}$ feature is an electronic Raman excitation -- a plasmon arising from the intrinsic hole self-doping widely reported in nominally undoped MnTe ($p \sim 10^{18}$--$10^{19}$~cm$^{-3}$ across independent Hall measurements) -- with a calculated plasma frequency and polarization selection rule matching experiment~\cite{Thapa2026}. Independently, a combined Raman/infrared/high-resolution XRD study has raised a third possibility: that the feature is not even intrinsic to $\alpha$-MnTe, reporting intensity highly irreproducible across sample positions and batches and tracking it instead to a secondary MnTe$_2$ impurity phase, while reassigning the intrinsic $E_{2g}$ mode to $\sim$100~cm$^{-1}$ and an infrared-active $E_{1u}$ mode to $\sim$155~cm$^{-1}$, with no evidence for broken inversion or six-fold symmetry within experimental resolution. Ref.~\cite{Uykur2026} does address this question experimentally and provides a path towards the solution, however, there are no stoichiometry controlled experiments so far. The field is thus left with three mutually incompatible explanations -- a genuine but exceptionally weak symmetry-broken phonon, a self-doping-induced plasmon, and an extrinsic secondary-phase artifact -- with no clear consensus, and resolving this carries direct consequences both for MnTe's space-group/spin-group classification and for interpreting any low-frequency magnon or chiral-magnon feature in the same spectral window.

\begin{figure*}[!ht]
\centering
\includegraphics[width=16cm]{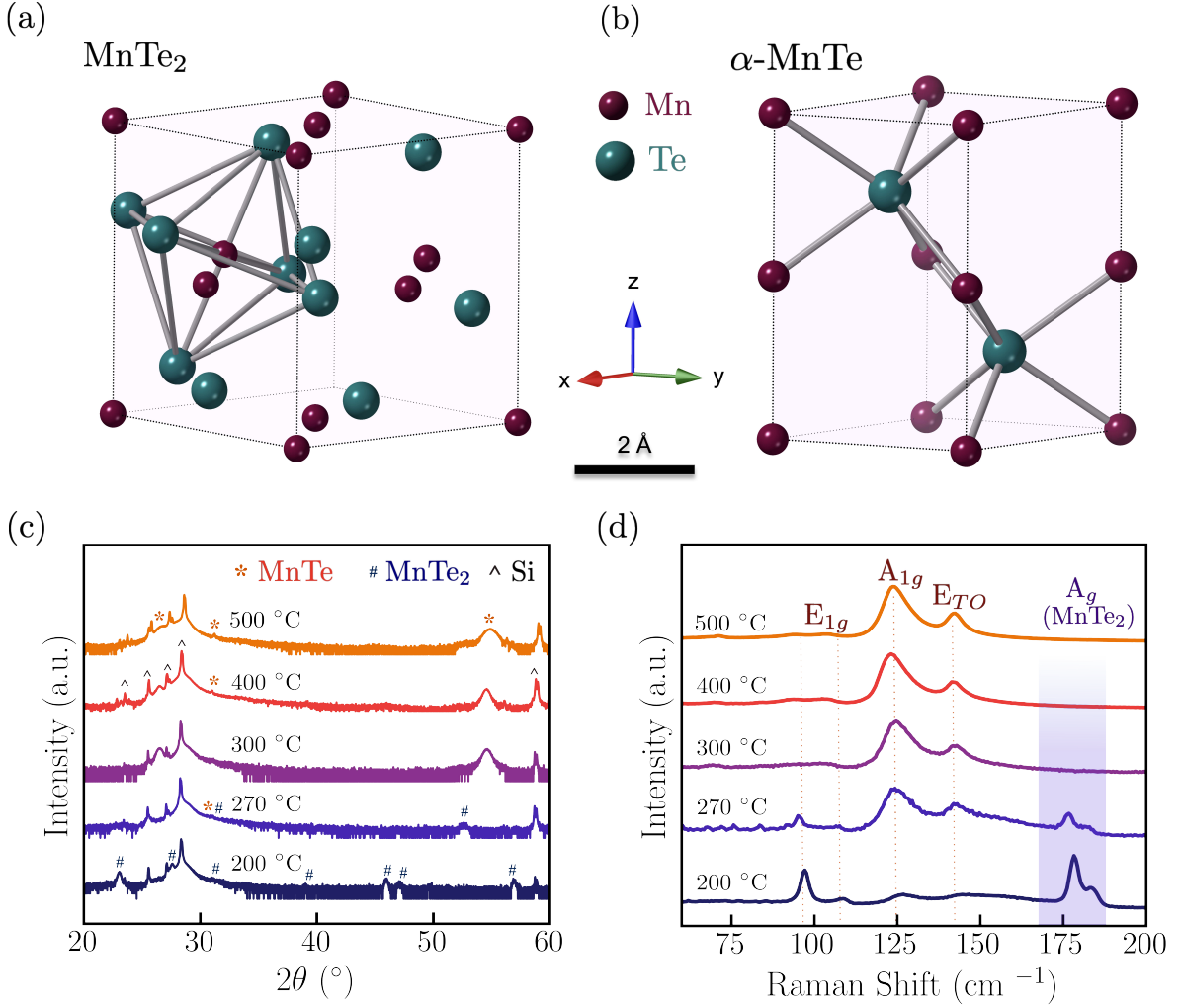}
\caption{The crystallographic arrangements of (a) MnTe$_2$ crystallize in pyrite cubic structure having space group Pa$\bar{3}$, where Mn is surrounding by six Te atoms forming octahedral MnTe${_6}$, while Te atoms form Te-Te dimers. (b) $\alpha$-MnTe crystallizes in hexagonal NiAs-type structure with space group $P6_3/mmc$, where Mn atoms occupy the octahedral sites between the Te layers. c) X-ray diffraction $2\theta$ scans for the MBE-grown MnTe series 
as a function of growth temperature (200--500\degree C). Peaks are indexed 
to the Si substrate ($\wedge$), $\alpha$-MnTe ($\ast$), and pyrite-type 
MnTe$_2$ ($\#$).  (d) Corresponding Raman 
spectra showing the intrinsic $E_{1g}$, $A_{1g}$, and $E_{TO}$ modes present 
across the series, alongside a feature near 175--180~cm$^{-1}$ (inset) assigned to $A_{g}$ of MnTe$_2$. }
\label{XRD and Raman}
\end{figure*}

In this work, we address this controversy from a purely materials perspective. Rather than relying on nominally single-phase bulk or solution-grown crystals, whose composition and defect chemistry are difficult to control precisely, we exploit the layer-by-layer stoichiometric control of molecular beam epitaxy (MBE) to grow a series of MnTe films spanning different growth temperatures  thereby varying the relative phase fraction of pyrite-type MnTe$_2$ against the target $\alpha$-MnTe matrix in a controlled, reproducible manner.  This growth-controlled correlation resolves the three-way ambiguity described above unambiguously in favor of the extrinsic MnTe$_2$ origin -- not through the spatial-inhomogeneity arguments of Ref.~\cite{Uykur2026}, but through deliberate, quantitative control of the impurity phase fraction itself.

Figure~\ref{RHEED} presents the RHEED patterns recorded across the growth 
series, beginning with the clean Si(111) substrate [Fig.~\ref{RHEED}(a)], 
which exhibits sharp, well-defined streaks characteristic of a well-ordered, 
reconstructed surface prior to MnTe deposition. Films grown at 200\degree C 
and 270\degree C [Figs.~\ref{RHEED}(b,c)] display diffuse patterns largely 
devoid of resolvable streaks or spots, indicating poor crystalline ordering 
consistent with insufficient adatom mobility at these low growth 
temperatures. At 300\degree C [Fig.~\ref{RHEED}(d)], the pattern sharpens 
into discrete, well-resolved spots rather than continuous streaks, the 
classic RHEED signature of a three-dimensional, island-mode growth front. 
At 400 and 500\degree C [Fig.~\ref{RHEED}(e,f)] does the pattern recover sharp, 
continuous streaks indicative of smooth, layer-by-layer two-dimensional 
growth. This temperature-dependent growth progression set forth a distinct structural trend: growth advances from an amorphous or poorly crystalline state at low temperatures, through a three-dimensional yet chemically ordered phase at intermediate temperatures, and culminates in optimal two-dimensional epitaxy only at the highest temperature examined, as shown by the morphological and topographical evolution added in SI Sections S1 and S2.

\begin{table}[t]
\centering
\renewcommand{\thetable}{1} 
\caption{Raman mode positions (cm$^{-1}$) as a function of growth temperature, observed in our samples.}
\label{tab:raman_modes}
\small
\setlength{\tabcolsep}{5pt}
\begin{tabular}{@{}llcccc@{}}
\toprule
Material & $T_{\mathrm{growth}}$ (\textdegree C) & E$_{1g}$ & A$_{1g}$ (Te) & E$_{TO}$ (Te) & A$_{g}$ \\
\midrule
\multirow{2}{*}{MnTe${_2}$}
 & 200 & 96, 108 & -- & -- & 178 \\
 & 270 & 95, 106  & 123 & 141 & 177 \\
\midrule
\multirow{3}{*}{MnTe}
 & 300 & 101 & 124 & 141 & -- \\
 & 400 & 94, 104 & 123 & 142 & -- \\
 & 500 & 93, 103 & 122 & 142 & -- \\
\bottomrule
\end{tabular}
\end{table}

Films grown at 200\degree C and 270\degree C 
show a dense set of reflections indexed to pyrite-type MnTe$_2$ in addition 
to the target $\alpha$-MnTe phase, confirming that the poor crystallinity 
observed by RHEED at these temperatures coincides with substantial 
secondary-phase formation~\cite{Martuza_2025}.  MnTe$_2$ adopts the pyrite structure (space group Pa$\bar{3}$), structurally distinct from the NiAs-type $\alpha$-MnTe matrix (Fig.~\ref{XRD and Raman}(a,b))~\cite{Xu_2018,Wu_2025_str}. This structural progression is corroborated and chemically resolved by XRD 
[Fig.~\ref{XRD and Raman}(c)]. 
At 300\degree C, the MnTe$_2$ reflections are 
largely suppressed and the pattern is dominated by $\alpha$-MnTe and Si 
substrate peaks, indicating that this film is chemically phase-pure despite 
its three-dimensional growth morphology~\cite{Zahra_2024}. The 400 and 500\degree C film is likewise 
dominated by $\alpha$-MnTe reflections with no significant MnTe$_2$ content, 
and is therefore the only growth condition that is simultaneously 
phase-pure and structurally optimal, i.e., grown in two-dimensional mode. 
The Raman spectra acquired for samples grown at different temperatures 
[Fig.~\ref{XRD and Raman}(d)],  reveal four features -- $E_{1g}$ 
($\sim$95--105~cm$^{-1}$), $A_{1g}$ ($\sim$125~cm$^{-1}$), $E_{TO}$ 
($\sim$145~cm$^{-1}$), and ($\sim$175--180~cm$^{-1}$) -- present at comparable relative intensity across various five growth conditions.  The highlighted feature near 175--180~cm$^{-1}$ confirms the MnTe$_2$ content determined independently by XRD: it is strong in the 200\degree C and appreciably weaker in 270\degree C films and essentially absent in the 300, 400, and 500\degree C films [Table 1]. The observed Raman shift of approximately 175 cm$^{-1}$ is ambiguous regarding its assignment to the E$_{2g}$ or A$_{g}$ mode. Experiments and theories have shown the  Raman shift in MnTe corresponding to 175 cm$^{-1}$  as the E$_{2g}$ mode \cite{Zhang2020}, whereas for MnTe$_{2}$ it is assigned to the A$_{g}$ mode \cite{Palasyuk2026}. However, our structural studies show that the sample grown at 200\degree C is indeed of MnTe$_{2}$ phase. Therefore, the Raman shift of approximately 175-180 cm$^{-1}$ is attributed to the A$_{g}$ mode of MnTe$_{2}$~\cite{Mller_1991}. Furthermore, the $A_{1g}$ and $E_{TO}$ at 270\degree C exhibit a remarkably broad signature. Above 270\degree C, the modes begin to form, which belong to isolated elemental Tellurium (Te)~\cite{ChowdeGowda_2025}, thereby corroborating the phase purity of MnTe$_{2}$ at 200\degree C. Notably, the weak Raman signature in the 300\degree C film -- despite its 3D growth morphology -- mirrors the XRD result above, confirming the mode tracks chemical phase content rather than surface morphology. The samples synthesized at 400\degree C and 500\degree C exhibit $E_{1g}$, $A_{1g}$, and $E_{TO}$ modes. Additionally, along with the absence of A$_{g}$ mode in these two samples  formation of MnTe phase can be inferred.

\begin{figure}[!ht]
\centering
\includegraphics[width=\columnwidth]{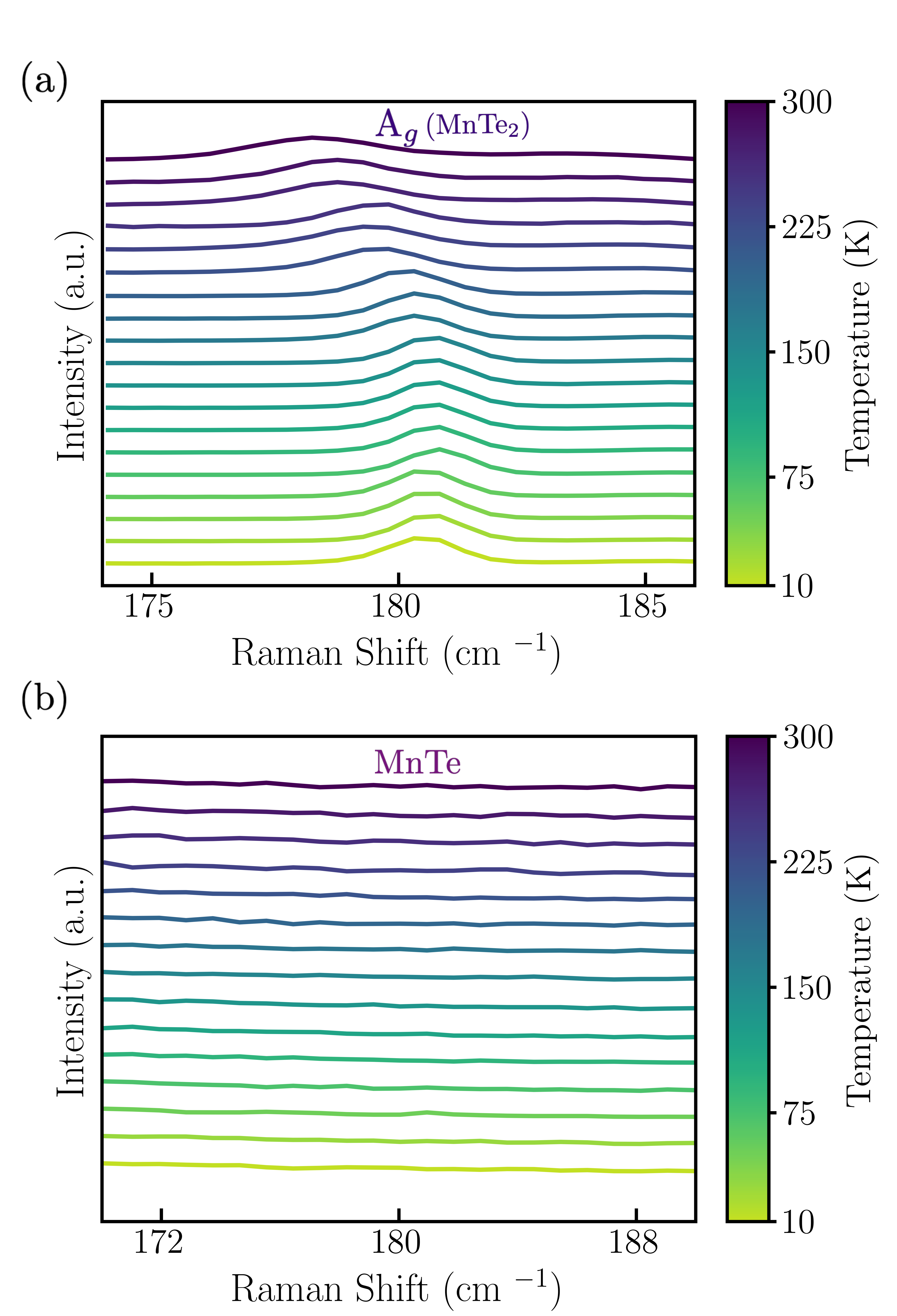} 
\caption{The temperature dependence Raman spectra of (a) MnTe$_2$, and (b) MnTe films.}
\label{Temp_raman}
\end{figure}

Finally, we investigate the temperature dependence of the Raman mode in the range 170~cm$^{-1}$ -- 190~cm$^{-1}$.  The Raman spectra of MnTe$_2$ [Fig.~\ref{Temp_raman}(a)] shows  temperature dependence of the peak width across the measured temperature range (300~K--10~K).  Notably, the 175 cm$^{-1}$ mode exists throughout the temperature range in the MnTe$_2$ film [see SI for analysis of peak width of the mode]. The visible shift in the peak position maybe attributed to minor change in strain due to lowering of temperature. Remarkably, the spectra of MnTe [Fig.~\ref{Temp_raman}(b)] shows no emergence of peak nor any significant temperature dependence. This distinct temperature-dependent behavior therefore serves as a spectroscopic fingerprint that clearly differentiates MnTe from MnTe$_2$.

\begin{figure}[ht]
\centering
\includegraphics[width=\columnwidth]{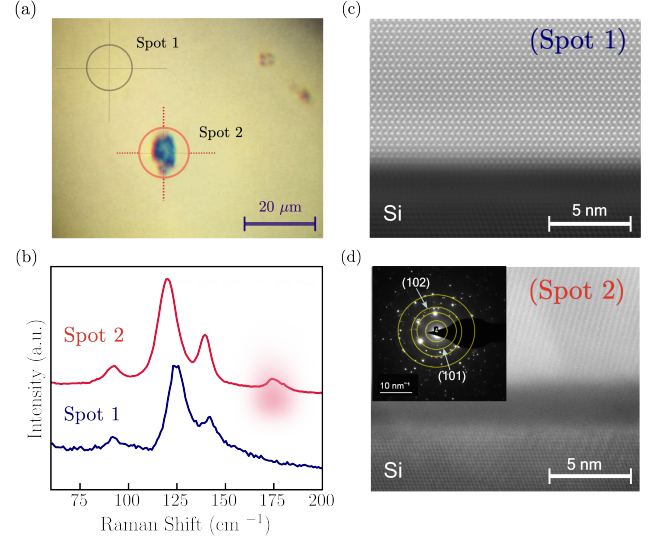} 
\caption{ (a) Optical microscope image acquired using Raman microscope of a surface of the $\alpha$-MnTe film; spot 1 shows pristine surface, while spot 2 shows a scorched surface due to increased Raman laser power and subsequent accumulations. (b) The Raman spectrum of the $\alpha$-MnTe film, measured at room temperature at spots 1 and 2. Transmission electron microscope images of (c) pristine region of the sample -- spot 1 (d) scorched region of the sample -- spot 2. Inset shows selected area electron diffraction over the same region.}
\label{scorched_raman}
\end{figure}

The presence of MnTe$_2$ is conclusively established by a characteristic Raman peak in the $175$--$180~\mathrm{cm}^{-1}$ range, attributable to the A$_{g}$  phonon mode \cite{Palasyuk2026}, accompanied by a shoulder peak displaced by $5$--$10~\mathrm{cm}^{-1}$ at room temperature. Likewise, the presence of MnTe is conclusively established by the appearance of Raman peaks at $95$--$105~\mathrm{cm}^{-1}$, $125~\mathrm{cm}^{-1}$, and $145~\mathrm{cm}^{-1}$. We increased the LASER power during Raman measurements, and found an additional mode at 175 cm$^{-1}$. Higher LASER power causes additional vibrational modes [Fig.~\ref{scorched_raman}(a, c)]. Beyond a threshold power, the region distorts. This suggests melting followed by recrystallization into polycrystalline phase with some segregation of Te. Notably, the distorted region exhibits the 175~cm$^{-1}$ mode, whereas the pristine region measured at nominal power does not. We therefore performed transmission electron microscopy on both the pristine and scorched regions of the same sample. The distorted region appears Mn-rich, suggesting Mn$_x$O$_y$ formation, and is poly-crystalline [Fig.~\ref{scorched_raman}(d)].

The 175~cm$^{-1}$ mode in this region [Fig.~\ref{scorched_raman}(b] is asymmetric. Deconvolution [see SI] yields two components, one of which we assign to an oxygen-rich Mn phase \cite{Son_2019,Okechukwu_2019,Bernardini_2025}. The other component persists at 175~cm$^{-1}$, which we attribute to distortions arising due to Te deficiency. Selected area electron diffraction obtained from the transmission electron microscope [Inset of Fig.~\ref{scorched_raman} (d)], exhibits planes of MnTe corresponding to the (101) and (102) planes of MnTe; the d-spacing obtained are 3.189~\AA~and 2.506 \AA~that align with the MnTe planes (101) -- 3.17 \AA~ and (102) -- 2.458~\AA~ respectively. A minor reduction in the lattice constant maybe attributed to LASER power impingement causing a compressive strain. This possibly triggers the observation of the Raman mode at 175 cm$^{-1}$. We can rule out the formation of  MnTe$_2$  in the scorched region. The temperature-dependent Raman spectra of the scorched region resemble those of pristine MnTe rather than MnTe$_2$, which further supports distorted MnTe. Determining the degree of distortion that produces this mode will require higher-precision, atomic-scale experiments. Our experiments establish that the mode corresponding to 175 cm$^{-1}$ is indeed MnTe mode; however, a careful examination is necessary before concluding the same because of proximity to the Raman mode corresponding to MnTe$_2$ as well. We can confirm with certainty that the observation of this mode corresponds to distorted phase rather than a pure epitaxial film, thereby the observation of this mode can serve as a tool to diagnose the crystalline quality of the samples.

Resolving the origin of the 175~cm$^{-1}$ mode through deliberate stoichiometric control, rather than further spectroscopic argument, suggests that some discrepancies accumulating in the MnTe literature may reflect sample-to-sample phase purity rather than exotic intrinsic physics. This distinction matters because MnTe continues to serve as the calibration standard against which new altermagnetic probes are benchmarked, making phase-purity verification as important as the measurement technique itself.   In this sense, the result functions less as an endpoint than as a tool for cleaning the spectroscopic record. With the 175~cm$^{-1}$ mode identified, genuine intrinsic features of $\alpha$-MnTe -- particularly low-frequency magnon and chiral-magnon excitations occupying the same spectral window can now be pursued with greater certainty.  As the field shifts from establishing altermagnetism as a phenomenon toward exploiting it in spintronic and magnonic devices, the reliability of underlying materials characterization will only grow more consequential. Extending this approach to other emerging altermagnetic candidates, especially thin-film systems prone to overlooked secondary-phase contamination, is a natural next step.

Authors are thankful to the National Nano-Fabrication Facility (NNFC) and the Micro and Nano Characterization Facility (MNCF) at the Centre for Nanoscience and Engineering, IISc. PB thanks Anusandhan National Research Foundation (ANRF), National Postdoctoral fellowship (PDF/2023/000444) for financial support.   DS thanks IISc start-up grant, Ministry of Electronics and Technology, Indian Space Research Organization, Infosys Foundation and Wadhwani Innovation Network for funding. Authors duly acknowledge funding from INOXCVA and INOX Airproducts for funding via CSR grants. 

\bibliography{abbreviation,references}

\end{document}


\renewcommand\thesection{\arabic{section}}
\renewcommand\thesubsection{\thesection.\arabic{subsection}}

\title{ Supplementary Text for \\Nature of the 175–180 cm$^{-1}$ Raman Feature in Altermagnetic $\alpha$-MnTe}

\author{Pankaj Bhardwaj}
\thanks{These authors contributed equally.}
\affiliation{Centre for Nanoscience and Engineering, Indian Institute of Science, Bengaluru, Karnataka 560012, India}

\author{Rajib Sarkar}
\thanks{These authors contributed equally.}
\affiliation{Centre for Nanoscience and Engineering, Indian Institute of Science, Bengaluru, Karnataka 560012, India}

\author{Naresh Shyaga}
\affiliation{Centre for Nanoscience and Engineering, Indian Institute of Science, Bengaluru, Karnataka 560012, India}

\author{Subhransu Kumar Negi}
\affiliation{Centre for Nanoscience and Engineering, Indian Institute of Science, Bengaluru, Karnataka 560012, India}

\author{Kartick Biswas}
\affiliation{Centre for Nanoscience and Engineering, Indian Institute of Science, Bengaluru, Karnataka 560012, India}

\author{Susmita Jana}
\affiliation{Condensed Matter Theory and Computational Lab, Department of Physics, IIT Madras, Chennai, 600036, Tamil Nadu, India}
\affiliation{Center for Atomistic Modelling and Materials Design, IIT Madras, Chennai, 600036, Tamil Nadu, India}

\author{Pavan Nukala}
\affiliation{Centre for Nanoscience and Engineering, Indian Institute of Science, Bengaluru, Karnataka 560012, India}

\author{B. R. K. Nanda}
\affiliation{Condensed Matter Theory and Computational Lab, Department of Physics, IIT Madras, Chennai, 600036, Tamil Nadu, India}
\affiliation{Center for Atomistic Modelling and Materials Design, IIT Madras, Chennai, 600036, Tamil Nadu, India}

\author{Dhavala Suri}
\email{dsuri@iisc.ac.in}
\affiliation{Centre for Nanoscience and Engineering, Indian Institute of Science, Bengaluru, Karnataka 560012, India}

\maketitle

\newpage
\renewcommand{\thesection}{S\arabic{section}}
\setcounter{section}{0}

\section{\NoCaseChange{Morphological Evolution of MnTe${_2}$ to $\alpha$-MnTe film}}

\begin{figure*}[!ht]
    \centering
    \includegraphics[width=\textwidth]{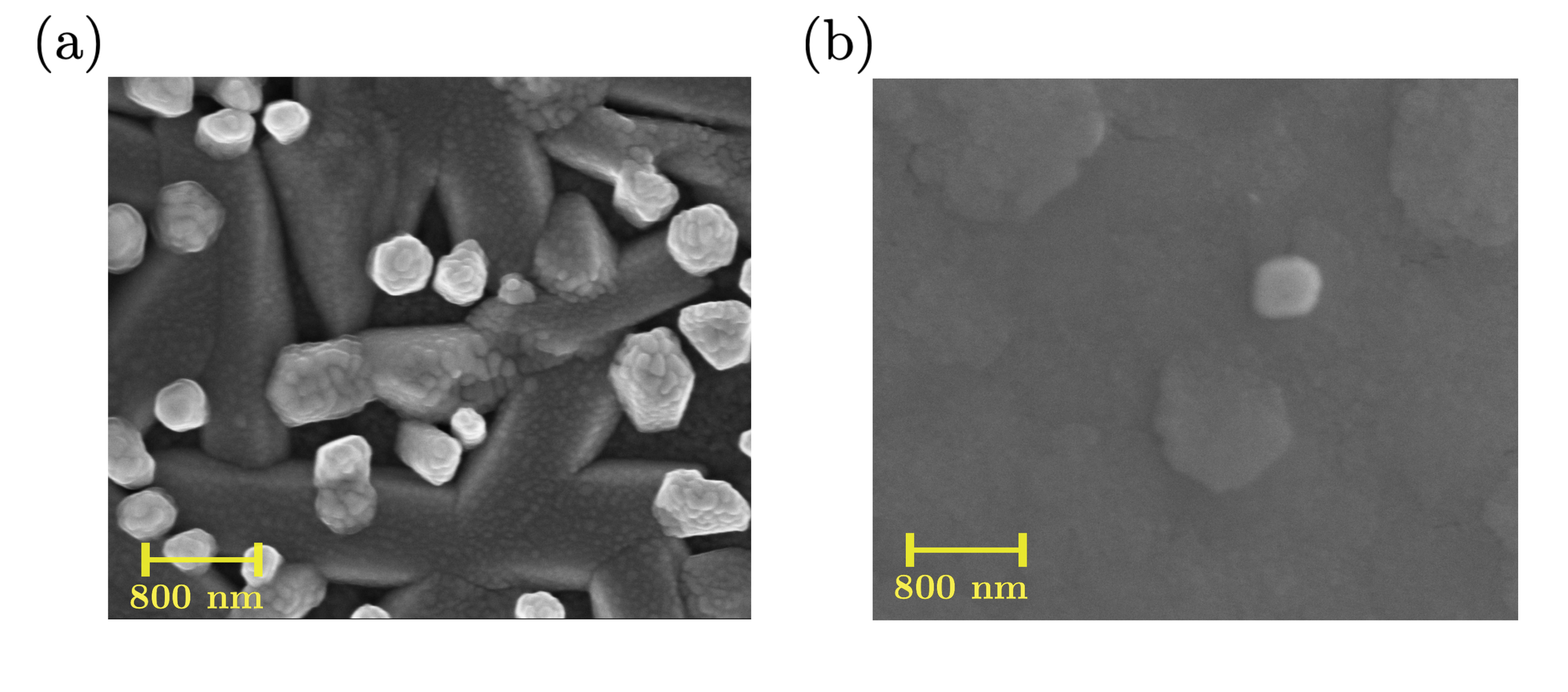} 
    \noindent \justifying Fig.~S1:~ Scanning electron microscopy of (a) MnTe${_2}$ ({200} {\celsius}) (b) $\alpha$-MnTe films ({400} {\celsius})
    \label{SI_SEM}
\end{figure*}

\noindent The morphological evolution of MnTe$_{2}$ to $\alpha$-MnTe thin films was investigated using Secondary Electron Microscopy (SEM). The SEM was an Ultra55 FE-SEM (Carl Zeiss) with EDS, operated at 100 X magnification and energy of 1.5 kV. The growth of {200}{\celsius} film results in the formation of pyrite-MnTe$_{2}$ phase, as confirmed by X-ray diffraction (XRD). Fig. S1 (a) displays large faceted cuboid plates and nucleated hexagonal nanoscales at grain boundaries, confirming columnar growth of MnTe$_{2}$ film. In contrast, {400} {\celsius} film results in the formation of NiAs - type hexagonal $\alpha$-MnTe phase. Fig. S1 (b) shows the disappearance of the coarse, faceted columnar cuboids and the observation of merged columnar grains and hexagonal facets, confirming that the secondary MnTe$_2$ phase is progressively eliminated and the formation of only $\alpha$-MnTe.

\clearpage

\renewcommand{\thesection}{S\arabic{section}}
\setcounter{section}{1}

\section{\NoCaseChange{Topographical Studies of MnTe${_2}$ to $\alpha$-MnTe film}}

\begin{figure*}[!ht]
    \centering
    \includegraphics[width=\textwidth]{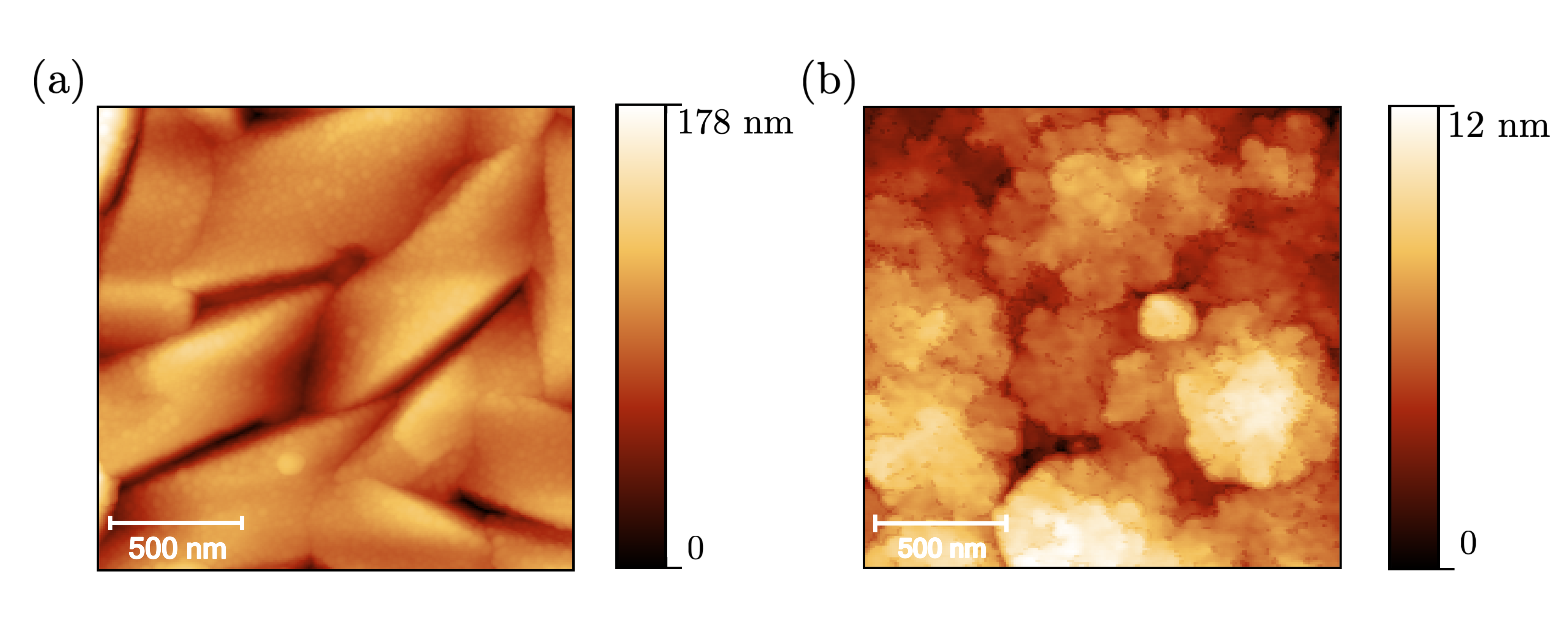} 
    \noindent \justifying Fig.~S2:~ Atomic force microscopy (AFM) of (a) MnTe${_2}$ ({200} {\celsius}) (b) $\alpha$-MnTe films ({400} {\celsius})
    \label{SI_AFM}
\end{figure*}
 
\noindent Topographic studies of MnTe$_{2}$ to $\alpha$-MnTe thin films were conducted using Atomic Force Microscopy (AFM) of the Bruker Dimension icon ICON AFM model, with a scan range of 2 $\times$ $\mu$m$^{2}$. Similar to SEM studies, Fig. S2 (a) depicts the coarsely columnar facets topology, characterized by a high height range of 178 nm and a root mean square roughness (R${_q}$) of ~18 nm. This observation aligns with the highly coarse, three-dimensional MnTe$_2$ grains inferred from the dense set of MnTe$_2$ reflections observed in XRD and SEM at growth temperature of {200} {\celsius}. Fig. S2 (b) illustrates the dense, fine-grained, granular texture, characterized by the smallest height range of 12 nm and the lowest surface roughness of 1.2 nm, confirming the homogeneous and smooth growth of $\alpha$-MnTe thin films.

\maketitle
\newpage
\renewcommand{\thesection}{S\arabic{section}}
\setcounter{section}{2}

\section{\NoCaseChange{Temperature-dependent Raman spectroscopy of the laser-induced thermally affected region in $\alpha$-MnTe film}}

\begin{figure*}[!ht]
    \centering
    \includegraphics[width=\textwidth]{Temp_Raman_MNte_burtn.pdf} 
    \noindent \justifying Fig.~S3:~ Temperature dependent Raman spectrum of (a) scroched surface of $\alpha$-MnTe films (b) Raman shift and FWHM of 175 cm$^{-1}$ (c) deocnvuluted Raman spectrum comprising peaks at 175 and 181 cm$^{-1}$.

    \label{SI_SEM}
\end{figure*}

\noindent Temperature-dependent Raman spectroscopy of the laser-scorched region of the MnTe surface reveals a mode at $\sim\!175~\mathrm{cm^{-1}}$, which has been attributed to the $A_g$ mode of a MnTe$_2$ phase~\cite{Thapa2026} or earlier claimed Raman-active $E_{2g}$ phonon of $\alpha$-MnTe~\cite{Osumi2024}. The mode is absent from the pristine surface and appears only in the scorched area. Atmospheric exposure can form a passivating Mn-oxide layer, which may contribute a feature near $171~\mathrm{cm^{-1}}$~\cite{Son_2019}, but a Lorentzian fit requires a second, weaker component on the high-wavenumber side of the main peak, so oxide alone does not account for the line shape. On warming from 5 to 290~K, the mode softens from $\approx 174.5$ to $\approx 177.5~\mathrm{cm^{-1}}$ and broadens from $\approx 3.5$ to $\approx 8.9~\mathrm{cm^{-1}}$, with a pronounced increase in linewidth and a drop in frequency above $\approx 240$~K. Such changes near a magnetic transition are consistent with spin--phonon coupling wherein FWHM changes correlate with spin fluctuations.~\cite{SS2025}. The anomaly appears below the bulk N\'eel temperature of $\alpha$-MnTe, which may be attributed to local laser heating or a modified magnetic response of the damaged layer.

\maketitle
\renewcommand{\thesection}{S\arabic{section}}
\setcounter{section}{3}

\section{\NoCaseChange{Comparison between Raman spectra of the pristine and scorched regions of MnTe, and MnTe${_2}$}}

\begin{figure*}[!ht]
    \centering
    \includegraphics[width=\textwidth]{SI_Temp_Raman.pdf} 
    \noindent \justifying Fig.~S4:~ Temperature dependent Raman shift and FWHM of 175-180 cm$^{-1}$ of (a) MnTe$_2$ (b) scroched region of $\alpha$-MnTe films.
    \label{SI_SEM}
\end{figure*}

\noindent For comparison, Fig S4 (a,b) shows the $\sim\!175~\mathrm{cm^{-1}}$ mode of MnTe${_2}$ and laser-induced thermally affected region in $\alpha$-MnTe film. On warming from 5 to 290~K, the MnTe$_2$ mode softens from $\approx 178.4$ to $\approx 181.5~\mathrm{cm^{-1}}$ and broadens from $\approx 1.5$ to $\approx 3.5~\mathrm{cm^{-1}}$, with a marked increase in linewidth and a decrease in frequency above $\approx 100$~K, which might be attributed to N\'eel temperature of MnTe$_2$ ($T_\mathrm{N}\approx 88$~K~\cite{Zhu2024}). In contrast, the mode in the laser-affected region of MnTe shows an anomaly in frequency and linewidth at $\approx 240$~K \cite{Sarkar_2026}, consistent with spin--phonon coupling. Because the two anomalies occur at different temperatures, and local laser heating would lower rather than raise the apparent transition temperature, eliminating the possibility of observed mode in the scorched region can be attributed to MnTe${_2}$ phase.

\begin{figure*}[!ht]
    \centering
    \includegraphics[width=\textwidth]{175MnTe_922.pdf} 
    \noindent \justifying Fig.~S5:~ Temperature dependent Raman shift and FWHM of 92-100 cm$^{-1}$ of (a) Pristine $\alpha$-MnTe surface (b) scroched surface of $\alpha$-MnTe, and (c) MnTe$_2$ films.
    \label{SI_SEM}
\end{figure*}
\noindent To identify the origin of the mode, we use the $\sim\!93$--$98~\mathrm{cm^{-1}}$ mode, which is present in MnTe$_2$ and in both the pristine and the laser-scorched regions of the $\alpha$-MnTe film, as a reference for the magnetic transition (Fig.~S5). Lorentzian fits show that this mode anomalizes near $100$~K in MnTe$_2$, consistent with its N\'eel temperature ($T_\mathrm{N}\approx 88$~K), whereas in both regions of MnTe the anomaly appears near $240$~K. The $175~\mathrm{cm^{-1}}$ mode of the scorched region shows the same $\approx 240$~K anomaly, which supports its association with the magnetic lattice of MnTe
rather than MnTe$_2$. We therefore assign the scorched region to a structurally distorted MnTe layer, in contrast to the MnTe$_2$ attribution. The onset near $240$~K, below the bulk value of $\approx 307$~K, is observed in pristine and scorched regions alike, and may reflect laser heating during the measurement or a reduced ordering temperature of the film.
\newpage
\begin{figure*}[!ht]
    \centering
    \includegraphics[width=\textwidth]{before_after_tem.pdf} 
    \noindent \justifying Fig.~S6:~ ~STEM-EDS (energy-dispersive X-ray spectroscopy) maps showing the distributions of Mn, and Te ions at (a) pristine $\alpha$-MnTe surface (b) scroched $\alpha$-MnTe surface
    \label{SI_SEM}
\end{figure*}

\noindent Scanning transmission electron microscopy with energy-dispersive X-ray spectroscopy (Fig.~S6) shows a reduced Te content in the scorched region relative to the pristine surface, together with a passivating Mn-oxide layer. A Te-deficient composition is incompatible with MnTe$_2$, which is Te-rich relative to MnTe, and is consistent with the absence of Raman signatures of elemental Te. We therefore describe the scorched region as Te-deficient, partially oxidized MnTe.

\newpage
\bibliography{references.bib}